\documentclass[journal abbreviation, manuscript]{copernicus}

\begin{document}
\nolinenumbers

\title{Towards GPROF V08: A Machine-Learning Framework for Simulating Passive Microwave Observations of Precipitation}

% \Author[affil]{given_name}{surname}

\Author[1][simon.pfreundschuh@colostate.edu]{Simon}{Pfreundschuh} %% correspondence author
\Author[1]{Paula J.}{Brown}
\Author[1]{Christian D.}{Kummerow}

\affil[1]{Department of Atmospheric Science, Colorado State University, Fort Collins, CO 80523, USA}

%% The [] brackets identify the author with the corresponding affiliation. 1, 2, 3, etc. should be inserted.

%% If an author is deceased, please add \deceased[$Deceased date if applicable$]{$Author number$} (e.g. \deceased[13 November 2015]{2}) at the end of the affiliations. The author number depends on the placement of the author in the author list, e.g. the third author has number 3.

%% If authors contributed equally, please add \equalcontrib{$Author numbers$} (e.g. \equalcontrib{1,3}) at the end of the affiliations. The author number depends on the placement of the author in the author list, e.g. the third author has number 3.

\runningtitle{Towards GPROF V08}

\runningauthor{Simon Pfreundschuh}

\received{}
\pubdiscuss{} %% only important for two-stage journals
\revised{}
\accepted{}
\published{}

%% These dates will be inserted by Copernicus Publications during the typesetting process.

\firstpage{1}

\maketitle

\begin{abstract}

 Spatially resolved precipitation retrievals based on convolutional neural
networks (CNNs) have demonstrated the potential to improve precipitation
estimates from passive microwave (PMW) observations and will be used to
implement the next operational PMW retrieval algorithm for the Global
Precipitation Measurement (GPM) mission. However, the conventional radiative transfer
simulations used to extend the retrieval algorithm to the various sensors of the
GPM constellation do not provide the spatial coverage required to train
CNN-based retrievals.

To overcome this limitation, we introduce a machine-learning-based simulation
framework that enables the training of the CNN-based GPROF-NN 3D retrieval
algorithm for the full GPM constellation. At the core of this framework is
Satformer, a new neural-network architecture that can translate PMW observations
between any pair of sensors of the GPM constellation. The resulting synthetic observations
exhibit lower errors than conventional radiative transfer simulations and lead
to more accurate precipitation retrievals when used to train GPROF-NN
precipitation retrievals. We further demonstrate the applicability of the
framework to new instruments by developing GPROF-NN retrievals for the Microwave
Imager aboard the  Weather System Follow-on Microwave satellite. While
GPROF-NN retrievals based on the new simulation framework achieve lower
instantaneous retrieval errors and higher correlations compared to independent
reference data than conventional retrievals, residual simulation errors can
introduce systematic biases in the resulting retrievals.

The proposed simulation framework provides a key component for extending
GPROF-NN to the full GPM constellation as part of the upcoming operational GPROF
V08 retrieval algorithm. The results demonstrate the potential of
machine-learning-based simulations to support spatially resolved precipitation
retrievals across heterogeneous satellite constellations.

\end{abstract}

\introduction  %% \introduction[modified heading if necessary]

Satellite-based precipitation measurements are essential for monitoring the
global distribution of precipitation, tracking the development and evolution of
storms and extreme precipitation events, as well as evaluating and ultimately
improving precipitation forecasts. The Global Precipitation Measurement (GPM)
mission \citep{hou_global_2014}, led jointly by NASA and JAXA, represents the
primary international effort to provide accurate, global precipitation estimates
at high spatial and temporal resolution. The mission relies on a constellation
of passive microwave (PMW) radiometers to maximize the spatial and temporal
coverage of precipitation-sensitive PMW observations.

The Goddard Profiling Algorithm (GPROF; \citealp{kummerow_evolution_2015}) has
served as the operational precipitation retrieval algorithm for all PMW sensors
of the GPM constellation since the start of the mission. Although the algorithm
has provided robust and reliable precipitation estimates, recent studies have
demonstrated that neural-network-based retrievals can improve the accuracy
\citep{pfreundschuh_gprof-nn_2022, pfreundschuh_gprof_2024} of PMW precipitation
retrievals. In particular the introduction of convolutional neural networks
(CNNs) that are able to leverage structural information in the satellite imagery
has led to notable advances in retrieval accuracy across a range of satellite
retrieval applications
\citep{pfreundschuh_improved_2022,gorooh_integrating_2023, amell2024,
amell_probabilistic_2025, pfreundschuh2026gprof}. Transitioning GPROF to a
CNN-based retrieval therefore represents a unique opportunity to improve
precipitation estimates from all current and historical PMW observations of the
GPM constellation.

However, \citet{pfreundschuh_gprof_2024} also identified several remaining
challenges that must be addressed to achieve consistent improvements in
precipitation estimates across all GPM sensors. The primary challenge is the
generation of training data for constellation sensors that do not overlap with
the Dual-Frequency Precipitation Radar (DPR), which provides the reference
precipitation estimates used to train the GPM PMW retrievals.

\begin{figure}[hbpt]
  \centering
  \includegraphics[width=\textwidth]{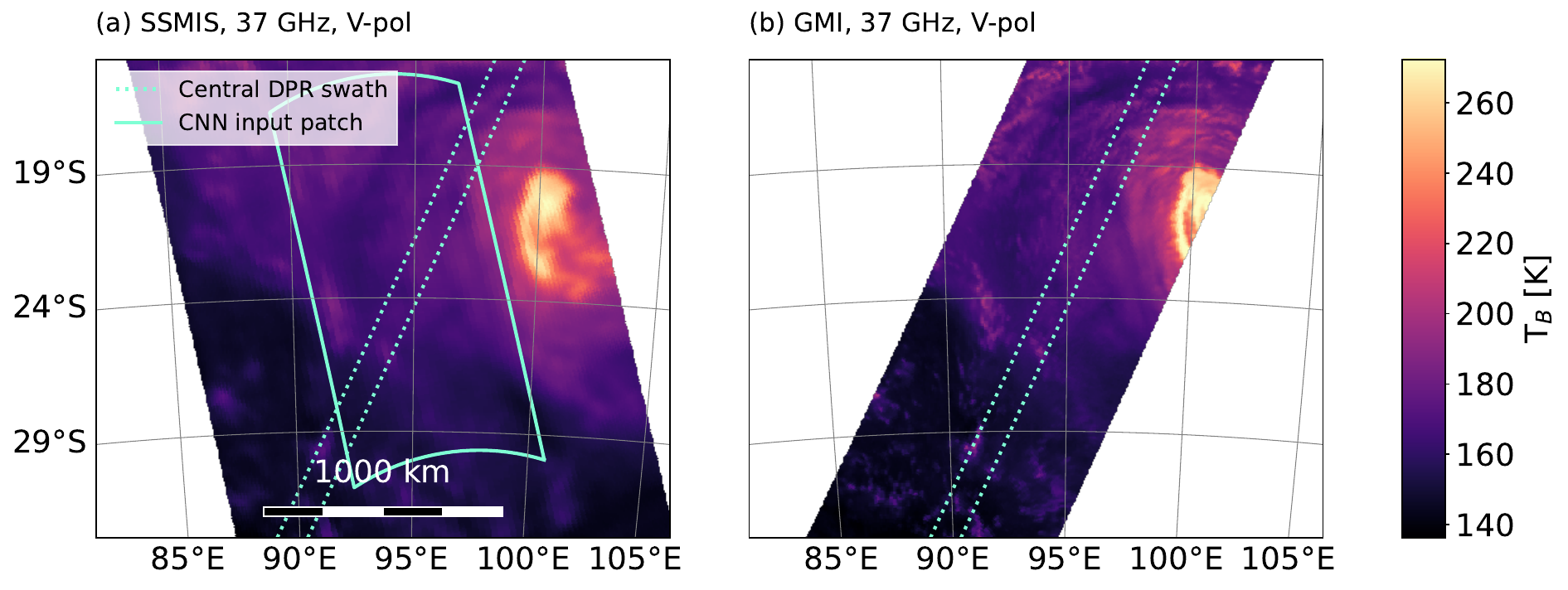}
  \caption{
    Comparison of actual SSMIS observations and the GPM-based data used to
produce the synthetic GPROF-NN training data for SSMIS. Panel (a) shows the 37
GHz vertically polarized observations from SSMIS. Panel (b) shows conincident 37
GHz, vertically polarized observations from GMI. The solid lines show the
spatial extent of a single image patch required to train the GPROF-NN retrieval.
The dotted line delineates the central part of the DPR swath for the
conventional radiative transfer framework can produce simulated SSMIS
observations.
  }\label{fig:obs_example}
\end{figure}

The conventional GPROF algorithm solved this problem using radiative transfer
simulations based on hydrometeor profiles retrieved from combined DPR and GMI
observations. Since these radiative transfer simulations require hydrometeor
profiles retrieved using the DPR, their spatial extent is limited to the DPR
swath. Since the conventional GPROF algorithm operated on all pixels
independently, the narrow swath of the DPR was sufficient to represent
observations from any GPM sensor. The spatial extent of the DPR swath, however,
is insufficient to fill the input-observation patches required to train the
GPROF-NN 3D retrieval. This limitation is illustrated in
Fig.~\ref{fig:obs_example}, which compares an input observation patch required by
the GPROF-NN 3D retrieval for the SSMI/S sensor with the corresponding portion
of the DPR swath for which simulated observations are available.

The first implementation of GPROF-NN 3D presented in
\citet{pfreundschuh_gprof-nn_2022} attempted to address this limitation by
training a neural network to emulate the simulated brightness temperatures used
by the conventional GPROF algorithm. However, the validation of GPROF-NN
retrievals for additional sensors presented in \citet{pfreundschuh_gprof_2024} demonstrated
that this approach was insufficient. In particular, errors in the simulated
observations were found to introduce substantial retrieval biases and, in some
cases, degrade retrieval accuracy relative to the conventional GPROF algorithm.

Adapting the convolutional GPROF-NN 3D retrieval to the full GPM constellation
 required the development of a new simulation framework capable of
generating accurate synthetic observations for all GPM constellation sensors.
Moreover, the framework must be able provide synthetic observations covering
the DPR swath from which the reference precipitation estimates are
derived and provide sufficient spatial coverage to cover image patches
spanning 128 scans and 64 pixels of the target sensor.

To address this challenge, we present the \textit{Sim V08} framework to generate
synthetic observations for all GPM PMW sensors. The framework is based on a novel
CNN architecture that enables sensor-agnostic translation of observations
between different GPM sensors. The sensor-agnostic design is a key feature of
the framework, which allows the simulator to be trained using collocations
between arbitrary pairs of GPM sensors. This capability is particularly
important for sensors that do not overlap with GMI, because it enables the
simulator to be trained indirectly through intermediate overlapping sensors.
Moreover, by relying primarily on PMW observations as input data, the framework
is capable of providing synthetic observations across the full GMI swath instead
of only the central DPR swath as was the case for the previous simulation framework.

This study is the first part of a two-part study describing the updated GPROF-NN
3D retrieval framework that has become the operational retrieval for the PMW
precipitation retrieval of the GPM constellation in early 2026 as \textit{GPROF
V08}. This first part describes the implementation of the new
neural-network-based simulation framework and provides an extensive validation
of both the simulated observations and the retrievals trained using them. A follow-up
study will describe final retrieval framework, which incorporates an
additional fine-tuning stage and evaluates the resulting retrievals against
independent precipitation reference datasets.

\section{Methods and Data}

A principal challenge for the design of the conventional GPROF as well as the
novel GPROF-NN retrieval algorithms is the need to support a constellation of
PMW sensors with varying observation characteristics. The original GPROF
algorithm was therefore designed as a parametric retrieval framework that can be
adapted to any PMW sensor whose observations can be accurately simulated. GPROF
traditionally used a shared database consisting of reference surface
precipitation estimates and hydrometeor profiles. Radiative transfer simulations
were then used to generate synthetic observations for each constellation sensor.
The precipitation estimates and simulated observations formed the
sensor-specific retrieval database, based on which the retrievals for the target
sensor were performed. In this way, the retrieval framework was able to support
a heterogeneous constellation of PMW sensors while enforcing consistent a priori
distributions across all sensors.

\subsection{The GPROF V07 Observation Simulation Framework}

The simulation framework used by GPROF V07 \citep{randel2020} -- denoted from
here on as \textit{Sim V07} -- uses conventional radiative transfer simulations
based on the Eddington approximation \citep{kummerow93} adapted for slant-path
simulations \citep{olson2001melting}. The simulations use hydrometeor profiles from the GPM Combined
Radar–Radiometer retrievals (GPM CMB; \citealp{grecu_gpm_2016}), complemented
with retrieved profiles from the Microwave Integrated Retrieval System
\citep{boukabara_mirs_2011} and atmospheric background fields from the JMA
Global Analysis System and surface emissivity estimates following
\citep{munchak2020}. The simulations are produced at the native 5 km $\times$ 5
km resolution of the CMB data and subsequently convolved with the sensor antenna
footprint to account for the varying spatial resolutions of different sensors.

\subsection{The GPROF V08 Observation Simulation Framework}

Since the Bayesian scheme used in GPROF V07 operated on single pixels, it was
sufficient for the Sim V07 framework to simulate independent pixels across the
central DPR swath. However, since the GPROF-NN 3D retrieval requires patches of
observations, the simulations for GPROF V08 must cover a sufficiently large
fraction of the input patch for each sensor, truthfully represent spatial structures, and
reproduce the footprint sampling and resolution of the target sensor. However,
as illustrated in Fig.~\ref{fig:obs_example}, the Sim V07 simulations cover only a
small fraction of the image patches required for the GPROF-NN 3D training.

To address these shortcomings and enable the training of the GPROF-NN 3D
retrieval across the full GPM constellation, we propose the SIM V08 framework.
It consists of two principal components: (1) a neural-network based simulator
that translates GMI observations into observations of any other GPM sensor and
(2) a sensor-footprint interpolator for mapping the observations from the GMI
swath to the spatial sampling of the target sensor. Since the simulator uses PMW
observations as principal input, it can generate synthetic observations across
the full GMI swath allowing the simulations to fill a significant part of the
input patches for most GPM constellation sensors. Additionally, the underlying
simulation model is designed to be sensor agnostic. Because of this, the
simulator can be trained using coincident observations of any pair GPM
constellation sensors thus enabling it to be trained even for sensors that never
overlap with GMI.

\subsubsection{The Satformer Architecture}

To implement the PMW-observation translator, we propose a novel neural-network
architecture called \textit{Satformer}, which enables sensor-agnostic
translation of PMW observations between different sensors. Rather than treating
observations as multi-spectral images with a fixed set of channels, the
Satformer represents the input as a sequence of two-dimensional observation
layers together with their associated observation properties. This allows a
single model to be applied to sensors with varying numbers of channels and
different characteristics. The observation properties are used to both
distinguish the input channels and specify the characteristics of the target
observations to be simulated.

The observation properties provide a numerical description of the sensor
characteristics that determine the principal properties of a sensor's
observations. For each observations pixel, the observation properties
are represented using an eight-element vector holding

\begin{enumerate}
  \item the central channel frequency in GHz,
  \item the offset from the central frequency in GHz,
  \item the weighting of vertically polarized radiation,
  \item the antenna beam width in degree,
  \item the sensor altitude in m, and
  \item the zenith angle and sine and cosine components of the azimuth angle of the line of sight.
\end{enumerate}

The satformer input data consists of a sequence of observation layers and
corresponding observation properties. Each observation layer holds a
$128\times128$ image patch from a given input channel. The observation layer is
accompanied by a corresponding rank-three tensor that stores the eight
observation-property features for each pixel in the image patch.

The Satformer architecture is based on a conventional U-Net-style
\citep{ronneberger_u-net_2015} encoder-decoder design. The encoder consists of
multiple stages of convolutional blocks separated by downsampling layers.
Convolution operations are applied independently across each observation layer in
the input sequence. To enable the exchange of information between observation
layers, each convolution block is augmented with an attention layer. The
attention mechanism operates pixel-wise across all observation layers, allowing
the model to combine information across the channels of the input sensor. The
use of attention is critical because it naturally supports variable-length
sequences and therefore enables the model to process observations from different
sensors with differing channel configurations.

To distinguish observations from different channels, the observation properties
associated with each layer are encoded using a two-layer multi-layer-perceptron
and added to the corresponding observations. Ancillary inputs, including
geostationary infrared observations, environmental temperature and humidity,
wind fields, and surface properties, are incorporated as additional input
layers. To allow the network to differentiate between observation and ancillary
inputs, a one-hot encoding of the input type is appended along the feature
dimension of each layer.

At the end of each encoder stage, the network produces progressively downsampled
sequences of encoded observation layers. A separate multi-head cross-attention
layer, using a learnable, length-one query sequence (as used in the Perceiver
architecture introduced in \citet{jaegle2021perceiver}), is then applied to map the
variable-length sequence of encoded observations to a sensor-independent,
fixed-size latent representation of the observed atmospheric state.

The decoder reconstructs a single layer of target observations from this latent
atmospheric representation. Its input consists of the multi-scale latent
atmospheric features produced by the encoder together with the properties of the
desired output observations. The output observation properties are downsampled
to match the spatial scale of the latent representation at each stage of the
decoder. The coarsest scale serves as input to the first decoder stage, while
finer-scale representations are incorporated through U-Net-style skip
connections by concatenating them with the output of the preceding decoder
stage.

\begin{figure}[hbpt]
  \centering
  \includegraphics[width=\textwidth]{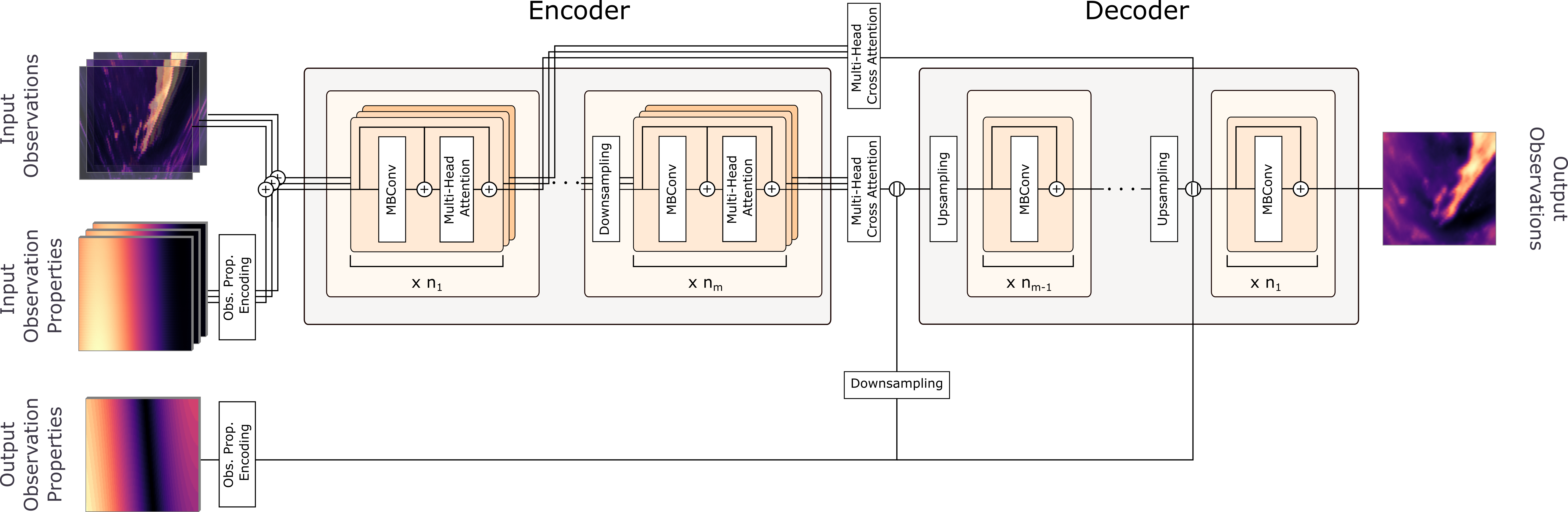}
  \caption{
    Illustration of the Satformer architecture used to translate observations
between different sensors of the GPM constellation. The model employs an
encoder–decoder architecture that maps a sequence of input observations from
multiple channels of a source sensor to observations from a single channel of a
target sensor. The encoder uses multi-head attention to exchange information
between channels. Cross-attention is used to map the variable-length sequence of
encoded input observations to a fixed-dimensional representation of the
atmospheric state. Finally, the decoder combines the encoded atmospheric
representation with a numerical description of the desired target-channel
properties to generate the simulated observations for the target sensor.
  }\label{fig:satformer}
\end{figure}

The resulting, complete Satformer architecture is illustrated in
Fig.~\ref{fig:satformer}. The detailed configuration is described in
Table~\ref{tab:satformer_details}.

\subsubsection{Training Data}

Because of its sensor-agnostic design, the training data for the simulator can
be generated from collocations between any pair of sensors in the GPM
constellation. Since the number of available collocations greatly
exceeds the available computational resources for training, the experiments
presented here use only collocations between TMI and GMI and the remaining
sensors in the GPM constellation. Because both TMI and GMI operate in precessing
orbits, they provide a large number of coincident overpasses with all other
constellation sensors.

From each collocation, training scenes of size 128 × 128 pixels are extracted
using the spatial sampling of the input sensor. To reduce variability between
scenes from different sensors, the input observations are linearly upsampled to
an approximate spatial resolution of $5\,\unit{km}\times 5\,\unit{km}$ at the
sub-satellite point. For each collocation between TMI or GMI and another
constellation sensor, training samples are generated in both directions: mapping
TMI/GMI observations to the constellation sensor observations, and mapping the
constellation sensor observations to TMI/GMI. The number of training
collocations used to train the Satformer are listed in table~\ref{tab:satformer_training_data}.

\begin{table}[t]
    \centering
    \caption{The number of Satformer training samples extracted from coincident observations of the GPM constellation.}
    \label{tab:satformer_training_data}
    \begin{tabular}{|c|c|c|c|}
        \hline
        Sensor Pair & Period Start & Period End & Number of Scenes \\
        \hline
        \hline
        GMI, AMSR2 & 2020-01-01 & 2021-01-01 & 39634 \\
        \hline
        GMI, ATMS & 2020-01-01 & 2021-01-01 & 108612 \\
        \hline
        GMI, MHS & 2020-01-01 & 2021-01-01 &  98546 \\
        \hline
        GMI, SSMIS & 2020-01-01 & 2021-01-01 &  84390 \\
        \hline
        TMI, AMSRE & 2008-01-01 & 2009-01-01 &  70207 \\
        \hline
        TMI, SSMI & 2008-01-01 & 2009-01-01 & 65638 \\
        \hline
        TMI, AMSUB & 2008-01-01 & 2009-01-01 & 58322 \\
        \hline
    \end{tabular}
\end{table}

\subsubsection{Training}

The simulator is trained simultaneously to reconstruct randomly removed input
channels and to predict observations from the target sensor. For each training
sample, the network predicts six output observation layers. Up to three of these
layers correspond to randomly dropped channels from the input observations,
while the remaining layers are randomly selected from the observations of the
target sensor. Training is performed for 120 epochs using the AdamW optimizer
\citep{loshchilov_decoupled_2019} with an initial learning rate of
$5 \times 10^{-4}$. The learning rate follows a cosine annealing schedule with
warm restarts \citep{loshchilov2017sgdrstochasticgradientdescent} after 20 and
60 epochs. Because the full training dataset exceeds 1 TB in size, the training
data are randomly subsampled by a factor of 10 at the beginning of each epoch.
The simulator is trained to predict 32 quantiles of the posterior distribution
of the output observations \citep{pfreundschuh_neural_2018}. This approach
enables both deterministic predictions and pixel-wise sampling of simulation
uncertainties.

\begin{figure}[hbpt]
  \centering
  \includegraphics[width=\textwidth]{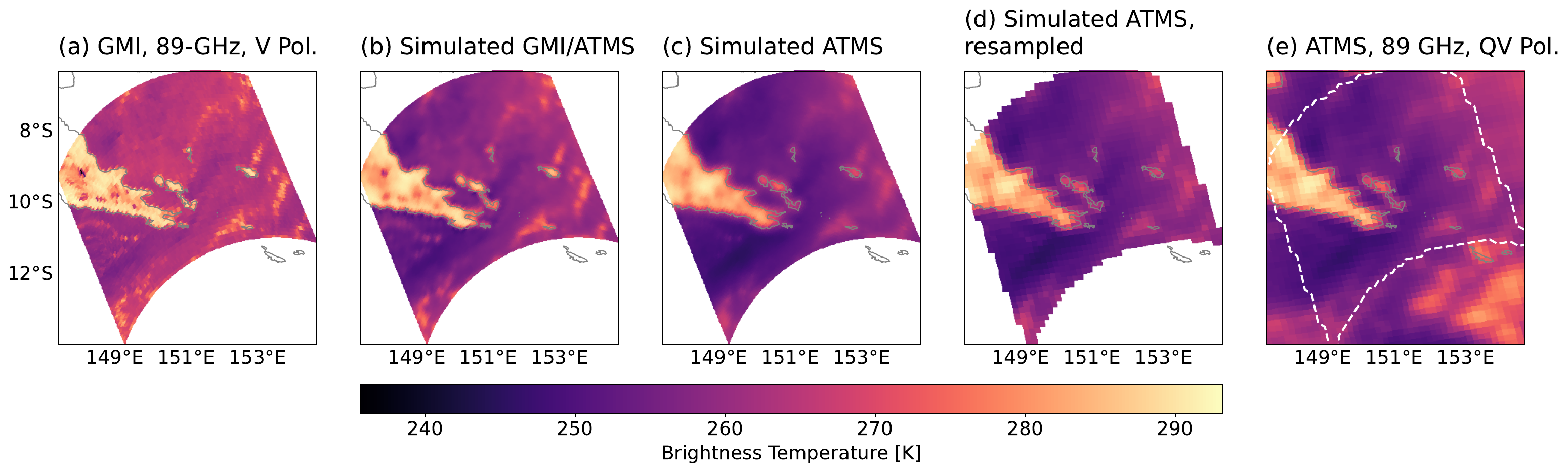}
  \caption{
Example of the Sim V08 simulator applied to translate GMI 89 GHz vertically
polarized observations into ATMS 89 GHz quasi-vertically polarized observations.
Panels (a) and (e) show the corresponding observed brightness temperatures from
GMI and ATMS, respectively. Panel (b) shows simulated observations generated
using output observation properties interpolated between those of the GMI and
ATMS channels. Panel (c) displays the simulated ATMS observations on the native
GMI swath, while panel (d) shows the simulations interpolated onto the ATMS
swath geometry.
  }\label{fig:satformer_example}
\end{figure}

Figure~\ref{fig:satformer_example} shows an example of the simulator
translating GMI observations into simulated ATMS observations. The figure shows
the (quasi-)vertically polarized 89 GHz channels from both GMI and ATMS.
Although the channels observe at nearly the same frequency, the resulting
observations differ substantially because of the distinct sensor
characteristics. Owing to its lower orbital altitude and larger antenna size,
GMI provides significantly smaller footprint sizes and therefore higher spatial
resolution. In addition, GMI is a conical-scanning radiometer that observes the
atmosphere at an approximately constant Earth-incidence angle with nearly true
vertical polarization. ATMS, in contrast, is a cross-track scanning sensor,
causing both the atmospheric path length and the effective polarization angle to
vary across the scan.

Because the output observation properties are provided explicitly as input
parameters, the simulator can continuously interpolate between the
characteristics of different sensors. Figure~\ref{fig:satformer_example}
demonstrates this capability by gradually varying the observation properties
from those of GMI to those of ATMS. Across the interpolation steps, the
simulator smoothly transforms the original GMI observations into realistic
ATMS-like observations. In particular, the spatial resolution decreases
progressively as the simulated sensor altitude and antenna beam widths are
increased toward the ATMS configuration.

When the Satformer-based synthetic observations are used to train GPROF-NN retrievals,
the predicted quantiles are used to generate pixel-wise independent random
samples from the posterior distribution of the simulated brightness temperatures
for each channel. Accounting for the uncertainties in the observations was
found to be crucial to produce accurate retrievals.

\subsubsection{Footprint Interpolator}
\label{sec:footprint_interpolator}

\begin{figure}[hbpt]
  \centering
  \includegraphics[width=\textwidth]{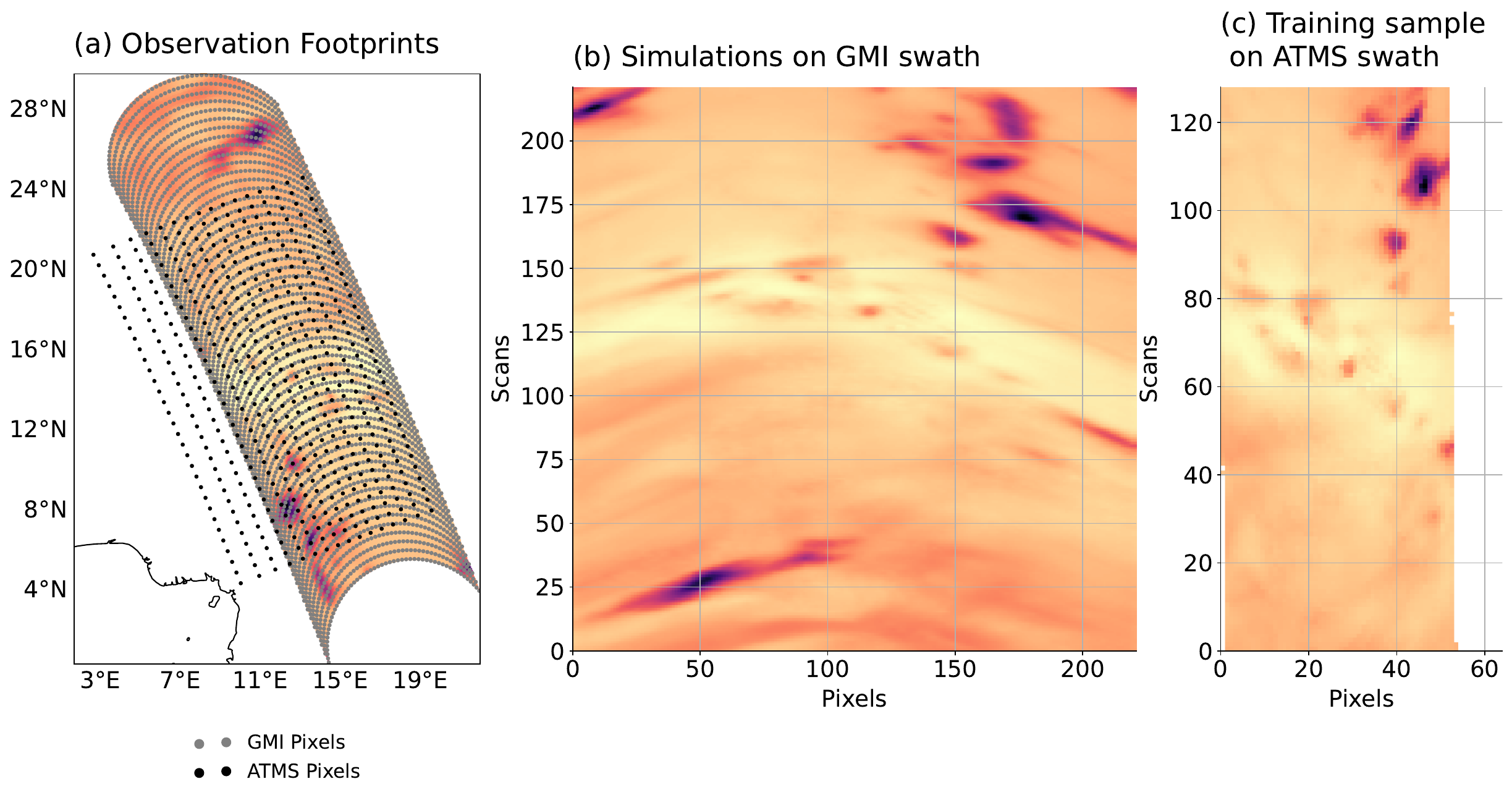}
  \caption{
    Example illustrating the remapping of simulated observations to the spatial
sampling of the target sensor. Panel (a) shows geolocated simulated observations
together with the locations of the GMI footprints (grey markers) and simulated
corresponding ATMS target pixels (black markers). Both GMI and ATMS pixels are
subsampled by a factor of four in along- and across-track directions to improve
 clarity. Panel (b) displays the simulated observations on the native GMI
swath geometry, while Panel (c) shows the same observations after remapping to
an image patch defined by the ATMS sampling geometry. }\label{fig:remapping}
\end{figure}

The Satformer model produces synthetic observations on the GMI swath. These
simulations are sufficient for training pixel-based retrievals such as GPROF-NN
1D. However, the appearance of geophysical features depends on the spatial
sampling of the observing sensor. Figure~\ref{fig:remapping} illustrates this
effect by comparing simulated observations on the native GMI swath (Panel b)
with the same observations remapped to the ATMS swath geometry (Panel c).

Because GPROF-NN 3D is designed to exploit spatial patterns in the observations,
producing realistic training data requires remapping the simulated GMI
observations to the spatial sampling geometry of the target sensor. To achieve
this, a target-sensor image patch is randomly selected from the target-sensor
swath and its locations relative to the GMI swath are calculated. The resulting
coordinates are then used to remap the GMI observations
onto the target sensor geometry using nearest-neighbor interpolation. This
remapping is performed on-the-fly during data loading. An example of the GMI
footprints and randomly calculated ATMS footprints is given in panel (a) of
Fig.~\ref{fig:remapping}. Although the GMI swath remains too narrow to fill the
full ATMS image patch, we found this to be sufficient for the training of the
GPROF-NN 3D retrievals. For cross-track scanning sensors, the Earth-incidence
angle varies across the scan. To account for this effect, simulations are
generated at ten discrete Earth-incidence angles. The final simulated
observations are then obtained by linearly interpolating between these
simulations using the Earth-incidence angle corresponding to each pixel in the
input observation patch.

\subsection{Validation Data}

The purpose of the proposed Sim V08 framework is the creation of training data for
the precipitation retrievals. To assess the accuracy of the resulting retrievals,
the following reference data will be used.

To validate retrievals over land, we will use gauge-corrected precipitation
estimates from the NOAA Multi-Radar Multi-Sensor (MRMS) product
\citep{smith_multi-radar_2016}. The radar-based reference precipitation and
satellite estimates are combined by first downsampling the MRMS data to a
resolution of 0.036$^\circ$ and then combining it with the satellite
precipitation estimates on a regular latitude-longitude grid with
a resolution of 0.036$^\circ$ following \citet{pfreundschuh2026benchmark}.
Low-quality radar estimates are removed by considering only pixels with an
average radar-quality index exceeding 0.8. The validation uses all CONUS
overpasses from 2022 for the four considered GPM sensors.

To validate the retrievals over the global oceans, we additionally use
collocations with the GPM combined radar and radiometer retrievals (2BCMB,
\citealp{cmb_data}). The 2BCMB reference data and PMW retrievals are collocated
by mapping them to the same 0.036$^\circ$-grid as the MRMS data. However, no
downsampling of the 2BCMB reference data is performed since its resolution is
around 5 km and thus lower than the grid resolution.

\subsection{Metrics}

For a given estimate $\hat{y}_i$ and corresponding reference value $y_i$ for
each sample $i = 1, \dots, N$ in the validation, we will use the relative bias, mean absolute error (MAE), mean-squared error (MSE), symmetric mean percentage error (SMAPE), and correlation coefficient to assess the accuracy of both the simulated observations as well as the
precipitation retrievals trained using the simulations. They are defined as

\begin{align}
  \label{eq:metrics_start}
  \mathrm{Bias} &= 100 \cdot \sum_{i=1}^{N}\frac{(\hat{y}_i-y_i)}{y_i} \\
  \mathrm{MAE} &= \frac{1}{N}\sum_{i=1}^{N}\left|\hat{y}_i-y_i\right| \\
  \mathrm{MSE} &= \frac{1}{N}\sum_{i=1}^{N}(\hat{y}_i-y_i)^2 \\
  \mathrm{SMAPE} &= \frac{1}{N}\sum_{i, 0.1 < y}^{N}\frac{|\hat{y}_i-y_i|}{0.5 \cdot (|\hat{y}_i| + |y_i|)} \\
  \text{Correlation Coef.} &= \frac{\sum_{i=1}^{N}(\hat{y}_i-\bar{\hat{y}})(y_i-\bar{y})}{\sqrt{\sum_{i=1}^{N}(\hat{y}_i-\bar{\hat{y}})^2 \sum_{i=1}^{N}(y_i-\bar{y})^2}}. \\
  \label{eq:metrics_end}
\end{align}

\section{Results}

Below we assess the performance of the proposed Sim V08 simulation framework in
multiple ways. First, we directly compare simulated observations from the
conventional Sim V07 and Sim V08 frameworks to actual observations. Second, we
assess the retrieval accuracy of GPROF-NN retrievals trained using the simulated
observations from both the conventional Sim V07 and the ML-based Sim V08
framework. Finally, we assess the capability of the ML framework to produce
training data for sensors not seen during training by assessing a GPROF-NN
retrieval for the  Microwave Imager \citep{draper2023} sensor on the Weather
System Follow-on - Microwave (WSF-M) satellite \citep{smith2025}.

\subsection{Simulation Accuracy}

\begin{table}[t]
    \centering
    \caption{Channel frequencies and polarizations of the
             SSMIS, AMSR2, ATMS, and MHS sensors as used by GPROF.}
    \label{tab:sensor_channels}
    \begin{tabular}{|r|cc|cc|cc|cc|}
        \hline
        & \multicolumn{2}{c|}{SSMIS}
        & \multicolumn{2}{c|}{AMSR2}
        & \multicolumn{2}{c|}{ATMS}
        & \multicolumn{2}{c|}{MHS} \\
        \hline
        Chan. & Freq. [GHz] & Pol. & Freq. [GHz] & Pol.
              & Freq. [GHz]& Pol. & Freq. [GHz] & Pol. \\
        \hline
        \hline
         1 & 19.35 & V & 10.65 & V & 88.2 & QV & 89.0 & QV \\
         2 & 19.35 & H & 10.65 & H & 165.5 & QH & 157.0 & QV \\
         3 & 22.24 & V & 18.7 & V & $183.31 \pm 1.0$ & QH & $183.31 \pm 1.0$ & QH \\
         4 & 37.0 & V & 18.7 & H & $183.31 \pm 3.0$ & QH & $183.31 \pm 3.0$ & QH \\
         5 & 37.0 & H & 23.8 & V & $183.31 \pm 7.0$ & QH & 190.31 & QV \\
         6 & 91.7 & V & 23.8 & H & & & & \\
         7 & 91.7 & H & 36.5 & V & & & & \\
         8 & 150 & H & 36.5 & H & & & & \\
         9 & $183 \pm 1.0$ & H & 89.0 & V & & & & \\
        10 & $183 \pm 3.0$ & H & 89.0 & H & & & & \\
        11 & $183 \pm 6.06$ & H & & & & & & \\
        \hline
    \end{tabular}
\end{table}

To assess the accuracy of the Sim V08 simulator, we extracted collocations
between the principal currently operational GPM constellation sensors (SSMIS, AMSR2, MHS,
and ATMS) and simulations generated using both the Sim V07 and Sim V08
frameworks. For each sensor type, the analysis uses one year of collocations,
spanning October 2018 through September 2019. The channels considered for each sensor
and their characteristics are listed are table~\ref{tab:sensor_channels}. Note that
the evaluation only considers channels used by the GPROF and GPROF-NN retrievals and thus
excludes temperature-sounding channels around 50 and 118 GHz.

Figure 5 compares the error characteristics of the two simulation frameworks.
Because the GPROF retrievals traditionally use only a specific subset of
channels with strong sensitivity to precipitation, most target channels have a
corresponding channel on GMI with similar spectral characteristics. For these
channels, the figure also includes error statistics obtained by directly using
the corresponding GMI observations as a proxy for the target sensor
observations.

Across all evaluated sensors and channels, the Sim V08 framework produces more
accurate brightness temperature simulations than Sim V07. Although the biases of
the Sim V07 simulations generally remain below 2\%, the Sim V08 simulations
consistently achieve substantially lower biases across all sensors and channels.
In addition, the machine-learning-based Sim V08 framework systematically reduces
RMSE and increases the linear correlation with the target observations. The
largest relative improvements are found for the cross-track scanning sensors,
with Sim V08 reducing the RMSE of the MHS and ATMS 89-GHz channels by
approximately a factor of two relative to Sim V07.

It is also noteworthy that, for some channels, directly using the corresponding
GMI observations as a proxy for the target observations still results in lower
RMSE and higher correlation coefficients than the Sim V07 simulations, and in
some cases also the Sim V08 simulations. For Sim V08, this behavior is limited
to a small number of AMSR2 channels, whereas for Sim V07 it occurs for both
AMSR2 and SSMIS channels. At the same time, using the GMI observations also
produces larger biases than the Sim V08 framework.

Because the GMI observations are included as inputs to the Sim V08 framework,
the simulator should in principle be capable of producing simulations that are
more accurate than the raw GMI proxy observations. The fact that this is not
consistently achieved suggests that the current model capacity or training
procedure may still limit performance, indicating that future versions of the
simulator could benefit from more complex model architectures or extended
training.

\begin{figure}[hbpt]
  \centering
  \includegraphics[width=\textwidth]{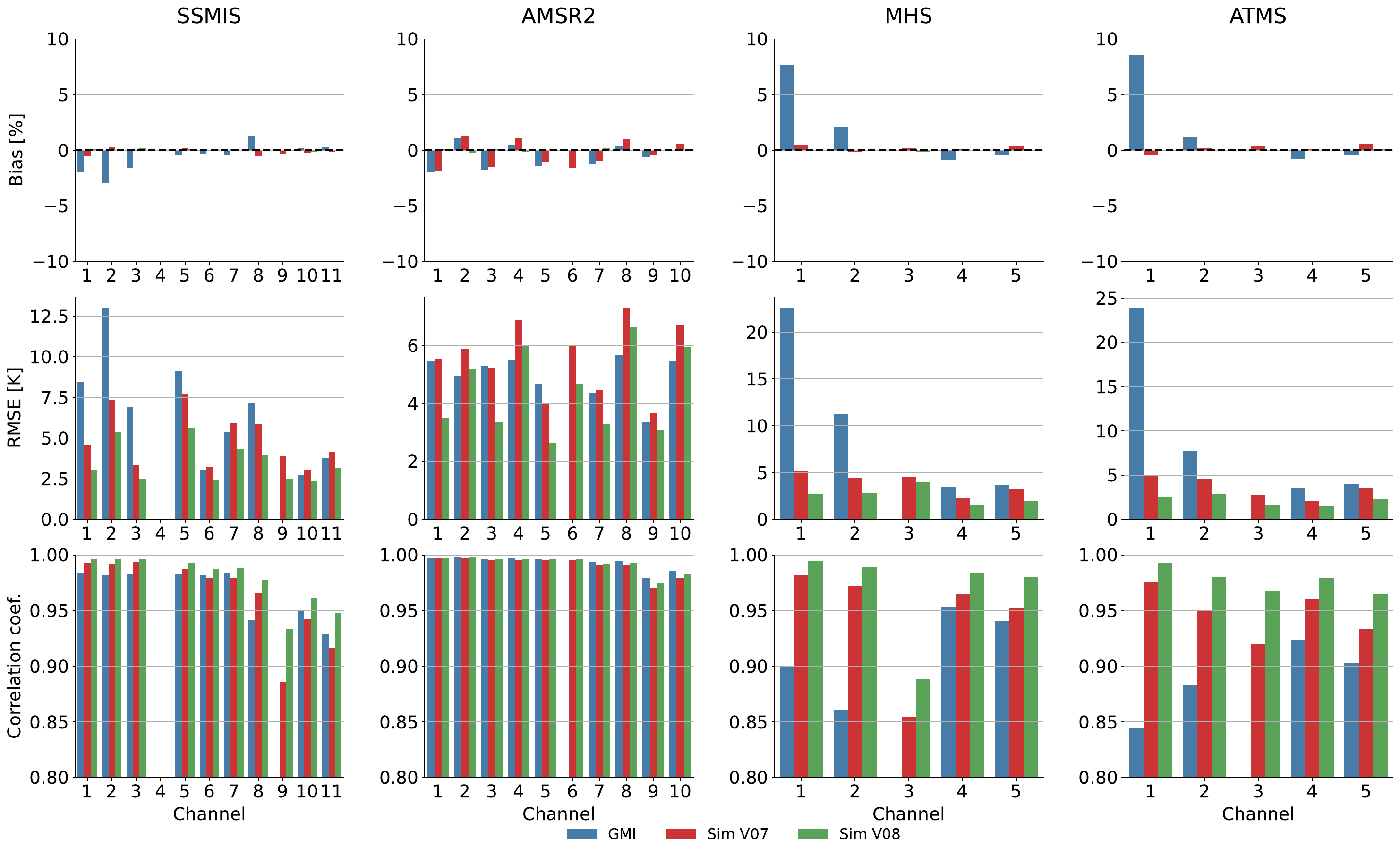}
  \caption{
          Comparison of error metrics for simulated observations generated using the
    Sim V07 and Sim V08 frameworks and baseline estimates derived from the
    corresponding GMI channels, where available. Separate columns show results for
    the principal operational GPM constellation sensors evaluated over the GPROF
    database period from October 2018 to September 2019. The rows show the relative
    bias, root-mean-square error (RMSE), and linear correlation coefficient,
    respectively.
  }\label{fig:simulator_accuracy}
\end{figure}

\subsection{Retrieval Accuracy}

Because the primary purpose of the simulated observations is to provide training
data for GPM precipitation retrievals, their most important characteristic is
their ability to produce accurate retrieval models. To assess this capability,
we compare retrievals trained using the Sim V07 and Sim V08 simulation
frameworks against independent precipitation estimates derived from ground-based
radar observations over CONUS and precipitation estimates from the 2BCMB product
over the global oceans.

To isolate the effects of the simulation framework from those of the spatial
remapping used for the CNN-based GPROF-NN 3D retrieval, we evaluate three
separate retrieval configurations for each sensor:

\begin{enumerate}
 \item GPROF-NN 1D (Sim V07): The pixel-based GPROF-NN 1D
\citep{pfreundschuh_gprof-nn_2022} retrieval trained using simulations generated
with the Sim V07 framework.
 \item GPROF-NN 1D (Sim V08): The pixel-based GPROF-NN 1D retrieval trained using simulations generated with the Sim V08 framework.
 \item GPROF-NN 3D (Sim V08): The image-patch-based GPROF-NN 3D retrieval trained using Sim V08 simulations together with the footprint interpolation  described in Section~\ref{sec:footprint_interpolator}.
\end{enumerate}

The climate version of the operational GPROF V07 retrieval is used as a
conventional baseline for comparison with these retrievals. The operational
GPROF V07 retrieval applies several post-processing corrections to both
simulated brightness temperatures and the retrieved precipitation. This includes
a cumulative distribution function matching applied to the input observations
and a surface-type-dependent bias correction. The latter is derived by adjusting
annual precipitation accumulations to match those of a reference sensor. In
contrast, the GPROF-NN retrievals evaluated here are trained directly on the
simulated observations and do not apply any corrections derived from
observational statistics.

\subsubsection{Case Study}

\begin{figure}[hbpt]
  \centering
  \includegraphics[width=\textwidth]{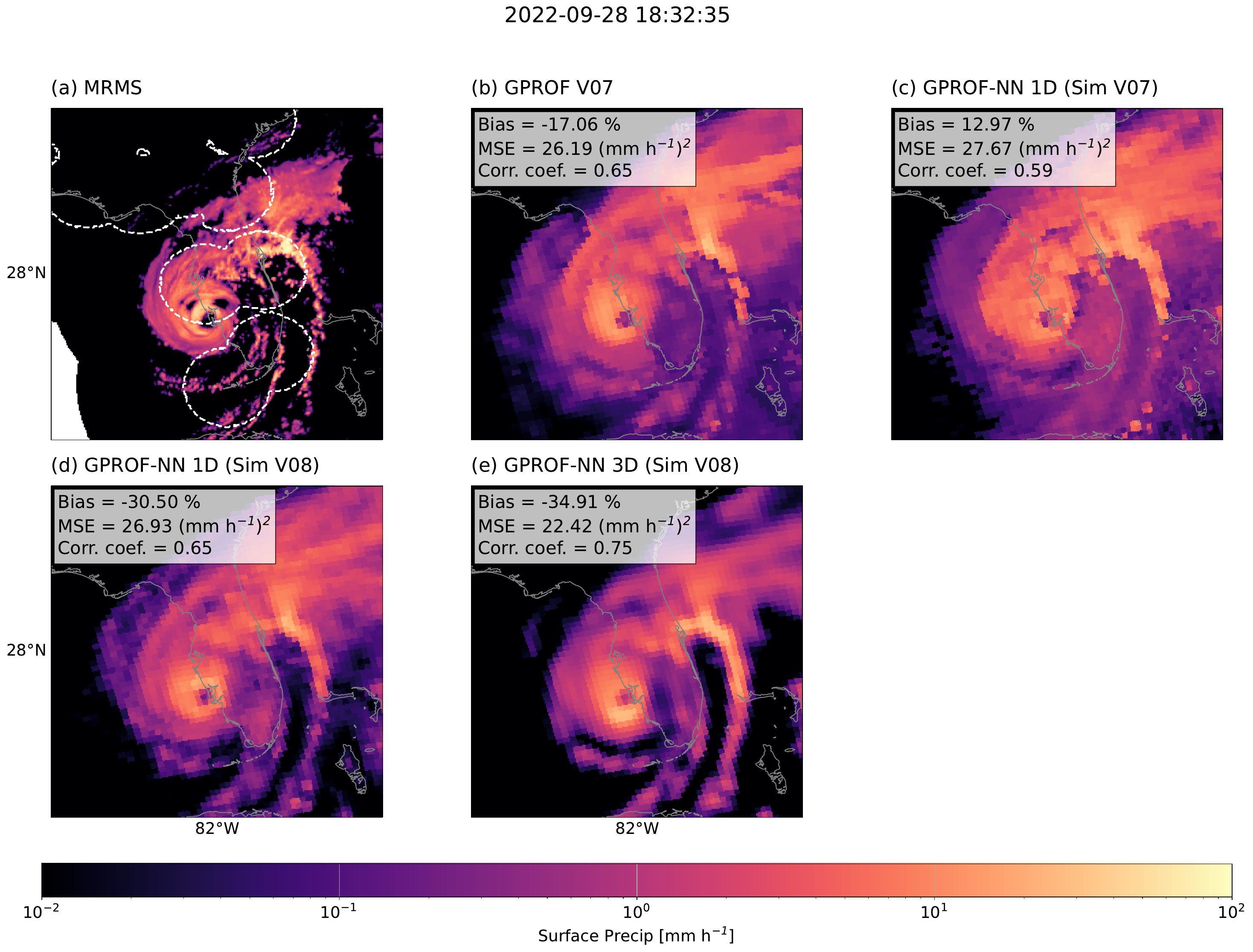}
  \caption{
    Case study of GPROF V07 and simulation-based GPROF-NN precipitation
retrievals from ATMS observations compared to MRMS reference estimates. Panel
(a) shows reference estimates from gauge-corrected ground-based precipitation
estimates from MRMS. White contours mark the regions with a radar-quality index
exceeding 0.8 used to calculate the error metrics. Remaining panels show the
results for (b) the GPROF V07 baseline, GPROF-NN 1D (Sim V07), (c) GPROF-NN 1D
(Sim V08), and (d) GPROF-NN 3D (Sim V08).
  }\label{fig:case_study}
\end{figure}

Figure~\ref{fig:case_study} compares precipitation estimates derived from ATMS
observations with reference precipitation from MRMS during the landfall of
Hurricane Ian on September 28, 2022. The GPROF V07 retrieval captures the
primary precipitation structures associated with the cyclone but substantially
overestimates the spatial extent of light precipitation south of the Florida
panhandle where MRMS data is reliable. The GPROF-NN 1D retrieval trained using
the conventional simulation framework produces results that are visually similar
to those from GPROF V07 but with a greater tendency to overestimate light
precipitation coverage. The GPROF-NN 1D retrieval trained using the Sim V08
framework reduces the amount of spurious light precipitation and yields moderate
improvements in the accuracy metrics relative to the retrievals based on the
conventional simulations. Visually, the GPROF-NN 3D retrieval trained using the
Sim V08 framework shows the closest agreement with the MRMS reference data. This
improvement is reflected in a reduction in mean squared error and an increase in
the correlation coefficient. The GPROF-NN 3D retrieval also exhibits the
strongest underestimation of total precipitation. However, since the retrieval
was not trained on MRMS it is unclear whether this reflects a shortcoming of the
retrieval itself or rather differences between the 2CMB-based reference data and
MRMS.

\subsubsection{Statistical Evaluation}

\begin{figure}[hbpt]
  \centering
  \includegraphics[width=\textwidth]{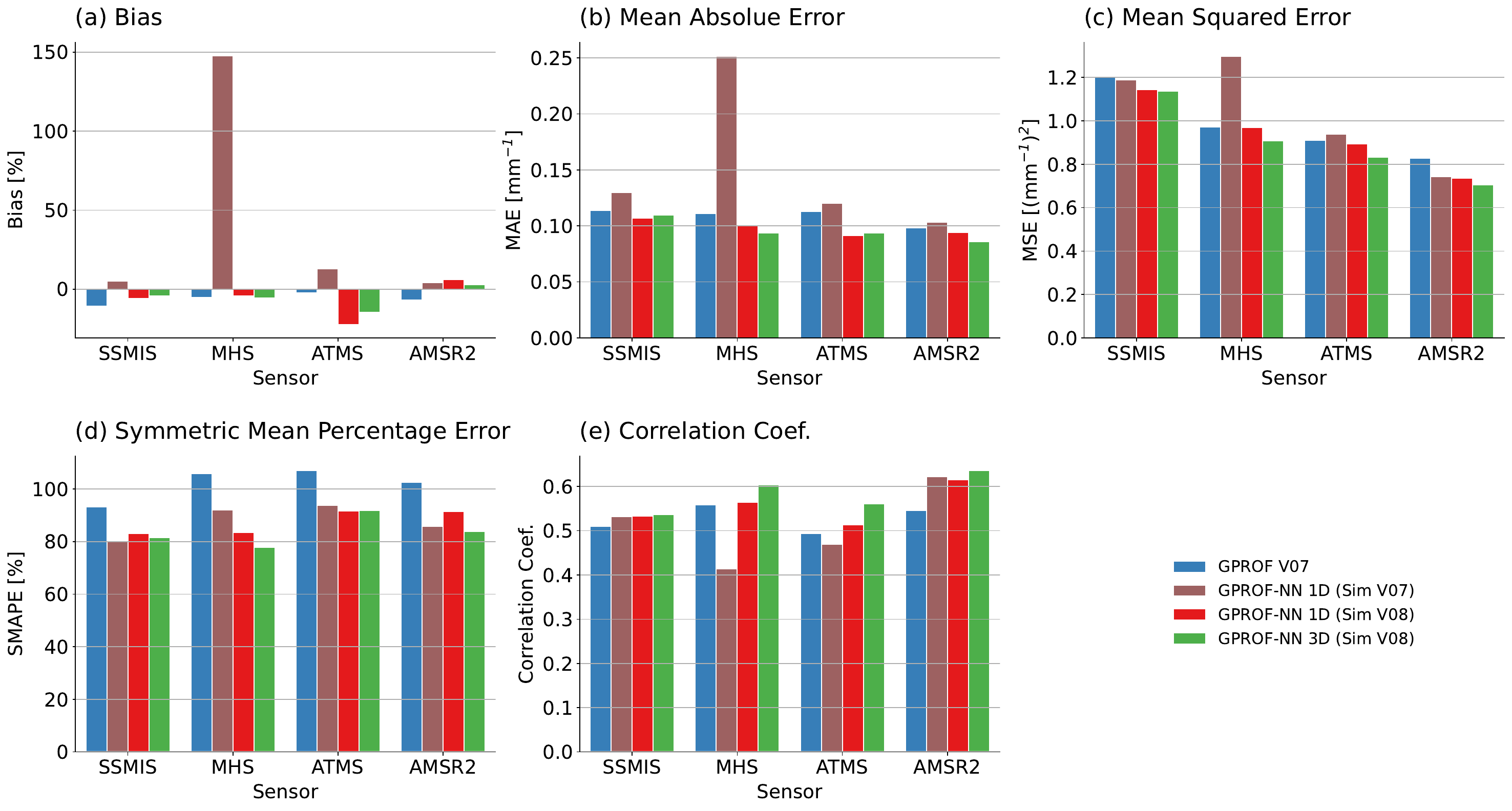}
  \caption{
    Retrieval accuracy metrics for the GPROF-NN retrievals trained using the Sim
V07 and Sim V08 simulation frameworks, together with the GPROF V07 baseline, for
the AMSR2, SSMIS, MHS, and ATMS sensors compared to ground-based reference
precipitation estimates from MRMS over CONUS. Panels (a)–(e) show the resulting
bias, mean absolute error, mean squared error, and correlation coefficient for
all retrieval configurations and sensors.
  }\label{fig:evaluation_conus}
\end{figure}

A systematic evaluation of the simulation-based retrievals compared to MRMS
reference estimates is presented in Figure~\ref{fig:evaluation_conus}. The
analysis is performed over all CONUS overpasses of AMSR2, SSMIS on F17, ATMS on
NOAA-20, and MHS on NOAA-19 sensors during the year 2022. For most retrievals,
the biases remain small. The exception is the GPROF-NN 1D retrieval trained
using the Sim V07 simulations, which exhibits a very strong positive bias and
high retrieval errors indicating a failure of the simulations to produce
accurate precipitation retrievals for this specific configuration. Notable
biases also occur for the ATMS retrievals. The persistence of retrieval biases
points to remaining inaccuracies in the simulations that the Sim V08 framework
is not able to address. However, for the comparison to the GPROF V07 baseline,
it must be kept in mind that the GPROF V07 retrieval applies both an empirical
correction to the input observations as well as a bias correction of the
retrieved precipitation rates. The GPROF-NN retrievals presented here are based
solely on simulations and can thus be applied from the first day of data
availability. Since such Day-1 retrievals are most relevant to real-time
applications, we deem these climatological biases acceptable. The operational
GPROF V08 retrieval will add an additional finetuning phase to the training as
well as a bias correct to remove these biases from long-term products. As will
be shown in the second part of this study, one year of observational data is
generally sufficient to remove remaining biases and further improve the
retrieval.

 In terms of instantaneous retrieval accuracy, both GPROF V07 and GPROF-NN 1D
trained using Sim-V07-based simulations show the weakest performance across the
evaluated sensors. Training the GPROF-NN 1D retrieval using the new Sim V08
simulation framework yields modest but consistent improvements over the GPROF
V07 baseline. The GPROF-NN 3D retrieval trained using the Sim V08 simulations
provides an additional increase in retrieval accuracy for all sensors
considered. The failure of the Sim V07-based GPROF-NN 1D retrieval to
consistently improve upon the GPROF V07 baseline highlights the challenge of
developing retrievals based on simulated observations, which may cause a more
sensitive retrieval method to degrade when trained on simulations but applied to
actual observations. The GPROF-NN 1D retrieval needs the more accurate
simulations from the Sim V08 framework to yield consistent improvements voer the
GPROF V07 retrieval. Further improvements in retrieval accuracy are achieved
when moving from the pixel-based GPROF-NN 1D to the image-patch-based GPROF-NN
3D retrieval. These additional improvements indicate that the Sim V08 framework
not only provides improved simulations but that the proposed footprint
interpolation scheme leads to an accurate representation of spatial features in
the simulated brightness temperatures.

The strong overestimation and increased retrieval errors observed for
GPROF-NN 1D (Sim V07) for MHS on NOAA-19  were traced to the MHS $183 \pm 1 \unit{GHz}$
channel. When this channel is dropped from the retrieval, the positive bias is
largely eliminated and the retrieval errors are reduced to levels consistent
with those observed for the other sensors. As illustrated in Fig. 8, the
$183 \pm 1 \unit{GHz}$ GHz channel on NOAA-19 MHS appears to be affected by
elevated noise levels during the evaluation period in 2022. Increased
root-mean-square differences relative to the corresponding ATMS channel are also
evident in Fig.~\ref{fig:simulator_accuracy}. The catastrophic degradation of
the GPROF-NN 1D (Sim V07) retrieval caused by the increased noise levels of the
$183 \pm 1 \unit{GHz}$ is representative of an increased sensitivity of the
neural-network-based GPROF-NN retrievals to observation artifacts and sensor
degradation.

\begin{figure}[hbpt]
  \centering
  \includegraphics[width=\textwidth]{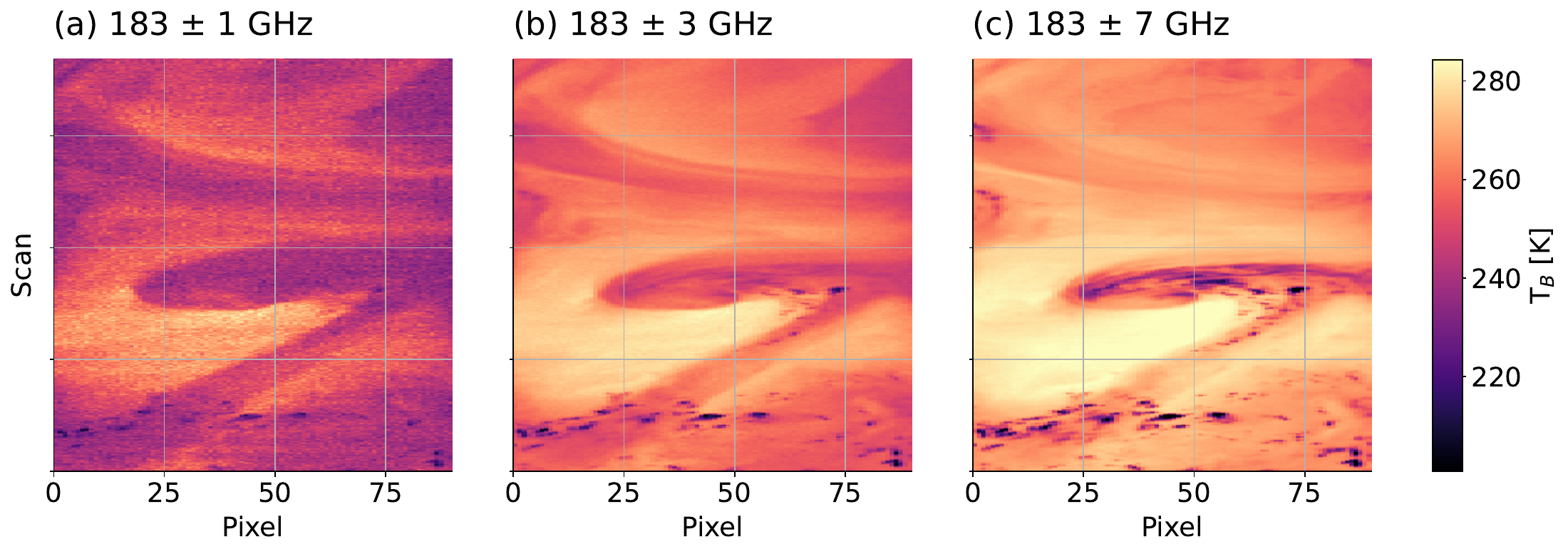}
  \caption{
    Comparison of the water vapor sounding channels of MHS on NOAA-19. The noise
    equivalent delta temperature of the $183 \pm 1 \unit{GHz}$ channel is $0.57\ \unit{K}$,
    which would not be visible on the color scale used. The visible noise
    suggests that the channel's noise levels are elevated causing the failure of the
    GPROF-NN 1D retrieval trained using Sim V07 based simulated observations.
  }\label{fig:mhs_noise}
\end{figure}

The Sim V08 simulation framework addresses observation uncertainty by sampling
from the predicted posterior distribution of the simulated observations. Because
the Sim V08 simulator was trained using NOAA-19 MHS observations, it learns to
reproduce the increased observation uncertainty associated with the degraded
channel. As a result, the GPROF-NN 1D (Sim V08) and GPROF-NN 3D (Sim V08)
retrievals remain more robust to the sensor degradation affecting NOAA-19 MHS.

\begin{figure}[hbpt]
  \centering
  \includegraphics[width=\textwidth]{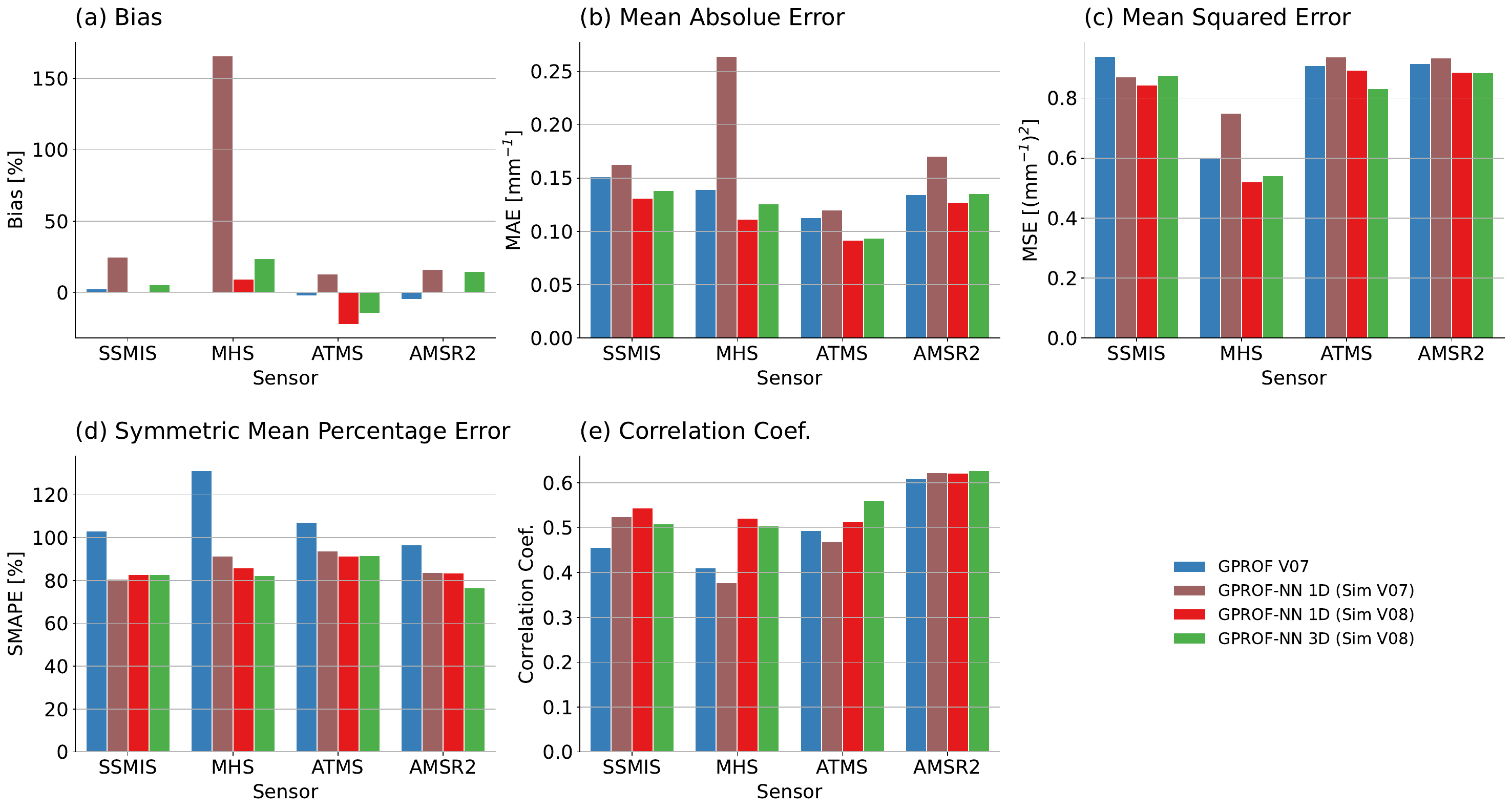}
  \caption{
    Like Fig.~\ref{fig:evaluation_conus} but evaluated against combined radar
    radiometer estimates from the 2BCMB dataset.
  }\label{fig:evaluation_ocean}
\end{figure}

Figure~\ref{fig:evaluation_ocean} evaluates the retrievals over oceans against
reference data from the 2BCMB dataset. The overall conclusions are consistent
with those obtained over land. Both the GPROF-NN 1D and GPROF-NN 3D retrievals
trained using the Sim V08 simulations provide robust  improvements
over the GPROF V07 baseline, further demonstrating the improved fidelity of the
Sim V08 simulation framework. However, the improvements are relatively smaller
than those observed over land. This is likely because reduced uncertainties in
surface emissivities lead to more accurate simulations and the additional
emission signal from precipitation in low-frequency channels lead to a higher
baseline accuracy.

It should be noted that collocations with the 2BCMB reference data occur most
frequently at high latitudes, causing the evaluation to be weighted toward
high-latitude precipitation regimes. Because precipitation at high latitudes
generally produces weaker PMW signatures and the DPR-based reference estimates
are subject to larger uncertainties, the resulting retrieval accuracies are
lower than those obtained in the evaluation over CONUS. In addition, the GPROF
V08 training dataset augments the 2BCMB-based reference data with light
precipitation and snowfall using the GPROF-NN XPR retrieval
\citep{gprof_nn_xpr}. This introduces discrepancies between the GPROF training
data and 2BCMB reference, which may decrease the overall accuracy of the
GPROF-NN retrievals.

\subsubsection{Application to WSF-M MWI}

The retrievals evaluated above were all based on sensors that the Sim V08
simulator has been explicitly trained on. An operationally more relevant
question is whether the Sim V08 framework generalized to observations from new
sensors. If the Sim V08 framework generalizes to novel sensors, it could be used
to develop day-1 retrievals that can be deployed on the first day of data
availability instead of waiting for a sufficiently large number of observations
to be available and retraining the simulator.

To evaluate this capability, we apply the Sim V08 framework to simulate
observations from the recently launched Microwave Imager (MWI,
\citealp{draper2023}) aboard the Weather System Follow-on Microwave 1
\citep{smith2025}. The channels used for the GPROF-NN retrieval are listed in
table~\ref{tab:gprof_nn_3d_channels}. MWI is a conically scanning PMW radiometer
with 18 channels spanning frequencies from 10 to 89 GHz. The GPROF retrieval
uses only the vertically and horizontally polarized channels and excludes the
23.8 GHz horizontally-polarized channel and the redundant 37.5 GHz channel. The
remaining channels are very similar to that of AMSR2, with the
primary difference being the substantially lower spatial resolution of the 89
GHz channels on MWI, which is approximately three times coarser than that of
AMSR2. Because of this similarity, the GPROF-NN 3D retrieval for AMSR2 can be
applied to MWI observations and is used here as a baseline for comparison.

\begin{table}[t]
    \centering
    \caption{Channels of the MWI and AMSR 2 used by GPROF.}
    \label{tab:gprof_nn_3d_channels}
    \begin{tabular}{|c|ccc|ccc|}
        \hline
        & \multicolumn{3}{c|}{MWI}
        & \multicolumn{3}{c|}{AMSR2} \\
        Chan. & Freq. & Pol. & IFOV [km $\times$ km]
              & Freq. & Pol. & IFOV [km $\times$ km] \\
        \hline
        \hline
        1 & 10.85 & V & $38 \times 23$
          & 10.65 & V & $42 \times 24$ \\
        2 & 10.85 & H & $38 \times 23$
          & 10.65 & H & $24 \times 42$ \\
        3 & 18.85 & V & $23 \times 15$
          & 18.7  & V & $22 \times 14$ \\
        4 & 18.85 & H & $23 \times 15$
          & 18.7  & H & $22 \times 14$ \\
        5 & 23.8  & V & $21 \times 13$
          & 23.8  & V & $19 \times 11$ \\
        6 &       &   &
          & 23.8  & H & $19 \times 11$ \\
        7 & 36.75 & V & $15 \times 10$
          & 36.5  & V & $12 \times 7$ \\
        8 & 36.75 & H & $15 \times 10$
          & 36.5  & H & $12 \times 7$ \\
        9 & 89.0  & H & $15 \times 10$
          & 89.0  & V & $5 \times 3$ \\
        10 & 89.0  & H & $15 \times 10$
          & 89.0  & H & $5 \times 3$ \\
        \hline
    \end{tabular}
\end{table}

\begin{figure}[hbpt]
  \centering
  \includegraphics[width=\textwidth]{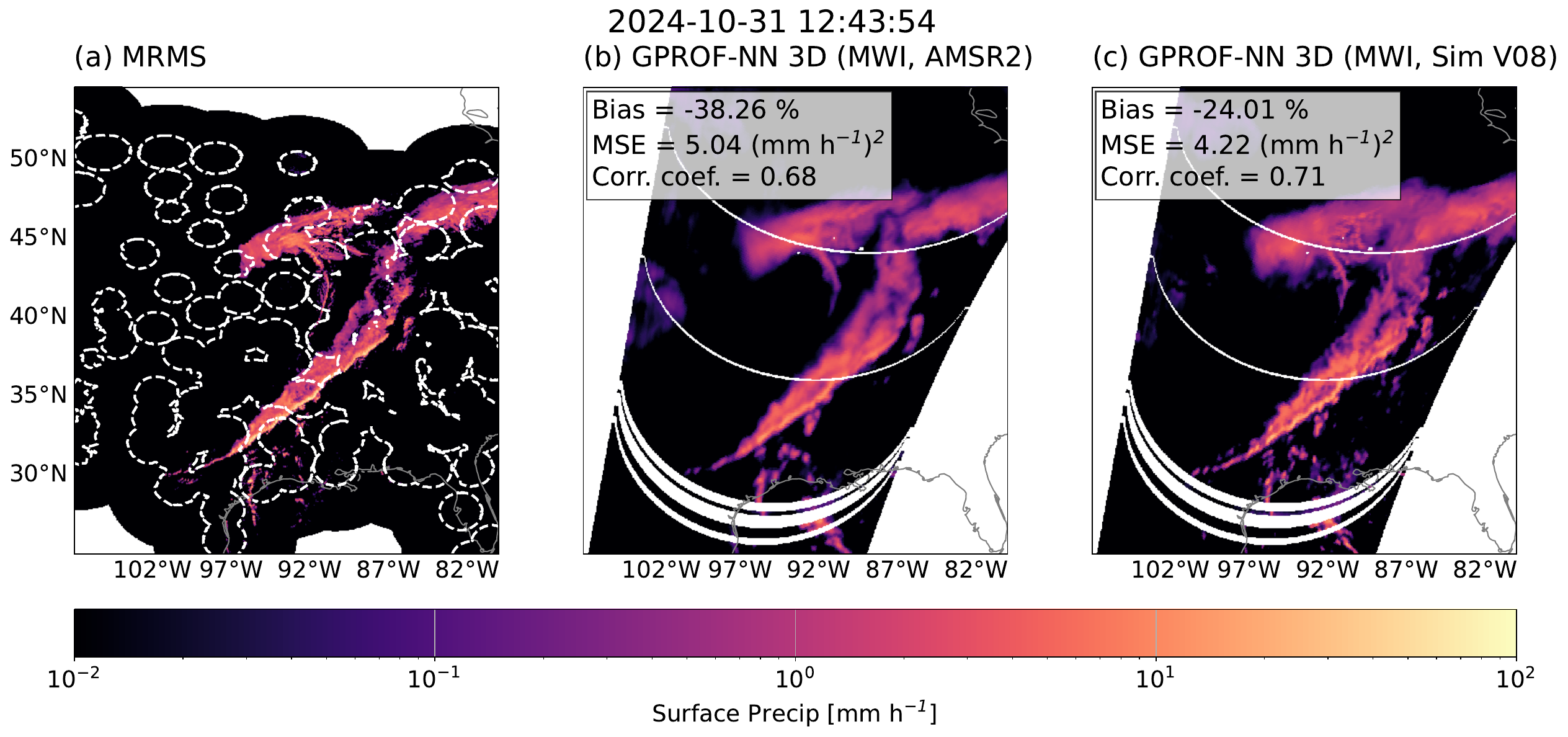}
  \caption{
    Comparison of GPROF-NN 3D day-1 retrievals for MWI against ground-based
precipitation estimates from MRMS. Panel (a) show gauge-corrected precipitation
estimates from MRMS. White contours mark the regions with a radar-quality index
exceeding 0.8 used to calculate the error metrics. Panel (b) shows results
obtained by applying the GPROF-NN 3D retrieval for AMSR2 to the MWI
observations. Panel (c) shows results of GPROF-NN 3D retrieval for MWI trained
using synthetic brightness temperatures from the Sim V08 framework.
}\label{fig:case_study_mwi}
\end{figure}

Figure~\ref{fig:case_study_mwi} compares retrieval results from the GPROF-NN 3D
retrieval for MWI trained using Sim V08 simulations to the results from the
operational GPROF V08 retrieval for AMSR2 applied to the MWI observations. The
operational GPROF V08 retrieval for AMSR2 is based on the same GPROF-NN 3D
retrieval but trained using Sim V08 simulations for AMSR2 and finetuned using
collocations of actual AMSR2 observations. Both retrievals successfully
reproduce the main precipitation structures associated with the storm system.
However, the MWI retrieval produces visibly sharper precipitation features and
more accurately separates the convective squall line from the trailing
stratiform precipitation. The improved resolution is reflected in a lower MSE
and higher linear correlation coefficient.

A systematic evaluation using MRMS overpasses from July through December 2024
confirms the findings from the case study. The results, summarized in Table 3,
shows modest improvements over the GPROF-NN 3D AMSR2 baseline across all
evaluated accuracy metrics except the mean absolute error. These results
demonstrate that the Sim V08 framework can generate high-quality synthetic
observations that capture the relevant observation characteristics required to
train spatially resolved precipitation retrievals even for previously unseen
sensors if their sensor characteristics are sufficiently close to those of
existing sensors.

\begin{table}[t]
    \centering
    \caption{
      Retrieval error metrics for the evaluations of MWI retrieval against MRMS
gauge-corrected precipitation estimates from June to September 2024.
           }
    \label{tab:mwi_mrms_metrics}
    \begin{tabular}{lccccc}
        \hline
        Retrieval & Bias [\%] & SMAPE [\%]
                  & MAE [mm h$^{-1}$]
                  & MSE [(mm h$^{-1}$)$^2$]
                  & Correlation Coef. \\
        \hline
        GPROF-NN 3D (MWI, AMSR2)
            & -25.54 & 87.22 & 0.11 & 1.4 & 0.59 \\
        GPROF-NN 3D (MWI, Sim V08)
            & -3.451 & 82.58 & 0.11 & 1.2 & 0.64 \\
        \hline
    \end{tabular}
\end{table}

\conclusions  %% \conclusions[modified heading if necessary]

This study presents a novel framework for generating synthetic PMW observations
for ML-based precipitation retrievals for the GPM constellation. The framework
was developed to overcome a key limitation of the conventional Sim V07
simulations, whose spatial coverage was restricted to the DPR swath because they
relied on vertically resolved hydrometeor retrievals from the 2BCMB product.
While sufficient for pixel-based retrievals, this limited coverage was
inadequate for the spatially resolved image patches required by the CNN-based
GPROF-NN 3D retrieval.

The newly developed Sim V08 framework addresses this limitation by using a
neural network to translate observations between different PMW sensors allowing
synthetic observations to be created across the full GMI swath. To enable this
capability, we introduced the Satformer architecture, which represents PMW
observations as sequences of observation layers characterized by their physical
observation properties rather than as fixed-channel multi-spectral images. This
sensor-agnostic design allows the simulator to be trained using collocations
between arbitrary pairs of overlapping sensors in the GPM constellation, making
it possible to support sensors that do not directly overlap with GMI.

A key feature of the Sim V08 framework is that the properties of the desired
output observations serve as inputs to the simulator model. This enables the
generation of observations for configurations that were not encountered during
training, provided their characteristics remain within the range of
characteristics represented in the training data. In addition to radiometric
properties such as frequency and polarization, the framework incorporates
antenna beam width, viewing geometry, and sensor altitude, allowing the
simulator to reproduce the spatial resolution of the target sensor. The
probabilistic output formulation further enables the simulation of realistic
observation uncertainty, including both thermal noise and simulation-related
uncertainty. With the footprint interpolation functionality described in
Section~\ref{sec:footprint_interpolator}, the Sim V08 functions as an end-to-end
PMW sensor simulator capable of generating physically realistic synthetic
observations for training CNN-based precipitation retrievals.

Direct validation against observed brightness temperatures demonstrated that the
Sim V08 consistently produces more accurate simulations than the conventional
Sim V07 framework. Retrievals trained using the Sim V08 simulations achieved
lower instantaneous errors and higher linear correlation than the GPROF V07
baseline.

The evaluation of the NOAA-19 MHS retrieval highlighted an additional advantage
of the ML-based simulation framework. Because the simulator is trained directly
on real observations and generates synthetic observations by sampling from a
probabilistic posterior distribution, the resulting retrievals are robust to
sensor artifacts and degradation. In particular, the framework successfully
handled the elevated noise levels affecting the NOAA-19 MHS $183.31 \pm 1.0$ GHz
channel, which caused catastrophic failures for retrievals trained using
simulations from the Sim V07 simulation framework.

Finally, we demonstrated that the Sim V08 framework can generalize to previously
unseen sensors. Retrieval experiments for the recently launched MWI sensor on
WSF-M showed that the framework can produce accurate day-1 retrievals without
requiring retraining of the simulator. This capability represents a notable
practical advance for operational precipitation retrievals, because it allows
retrieval algorithms to be deployed immediately after calibrated observations
become available. In contrast, the conventional Sim V07 framework typically
required at least a year of observations to derive the bias corrections and CDF
matching needed for operational use.

The development of the Sim V08 framework constitutes a critical step in the
operationalization of the GPROF-NN retrievals that will become the operational
GPM retrievals as GPROF V08. The framework provides more accurate
synthetic observations than the conventional simulation approach and enables
improved retrieval accuracy across the GPM constellation.

However, our evaluation also showed that the improved accuracy of the Sim V08
framework is insufficient to completely eliminate residual biases. Furthermore,
it was found that retrievals trained exclusively using synthetic observations
still do not fully match the performance achievable when training on actual
observations. For these reasons, the operational GPROF V08 implementation
introduces an additional fine-tuning stage using actual observations whenever
suitable reference data are available. The complete operational framework,
together with its evaluation against independent precipitation reference
datasets, is presented in the second part of this study.

\codedataavailability{

  The implementation of the Sim V08 framework is available through the
\texttt{gprof\_nn} Python package \citep{gprof_nn_github}. The GPM L1C
observations, GPROF V07 L2A baseline precipitation estimates, and 2BCMB
reference precipitation estimates are available from \citet{nasa_earthdata}.
MRMS precipitation estimates over CONUS are available from \citet{mrms_data}.

}

\appendix
\section{Satformer Architecture}    %% Appendix A

Table~\ref{tab:satformer_details} lists the detailed components of the Satformer model used
to implement the Sim V08 framework.

\begin{table}[hbpt]
    \centering
    \caption{Detailed architectural configuration of the Satformer model used to implement
      the Sim V08 framework.  The expansion factor $e$ specifies the channel expansion
      within the MBConv blocks.}
    \label{tab:satformer_details}
    \begin{tabular}{|l|c|c|c|c|c|}
        \hline
        \multicolumn{6}{|c|}{\textbf{Observation Property Encoding}} \\
        \hline
        Stage & Chan. In & Chan. Out & \multicolumn{3}{c|}{Conv. Block} \\
        \hline
        1 & 8 & 49 & \multicolumn{3}{c|}{3 $\times$ Linear, Layer Norm, GELU} \\
        \hline
        \multicolumn{6}{|c|}{\textbf{Stem}} \\
        \hline
        Stage & Chan. In & Chan. Out & \multicolumn{3}{c|}{Conv. Block} \\
        \hline
        1 & 1  & 49  & \multicolumn{3}{c|}{1 $\times$ Conv2d}  \\
        \hline
        \multicolumn{6}{|c|}{\textbf{Encoder}} \\
        \hline
        Stage & Chan. In & Chan. Out & Conv. Block & Down-/Up-sampling & Att. Heads \\
        \hline
        1 & 49  & 49  & 2 $\times$ MBConv, e=1, GELU, Layer Norm & - & \\
        2 & 49  & 64  & 3 $\times$ MBConv, e=2, GELU, Layer Norm & 2 & 4 \\
        3 & 64  & 96  & 3 $\times$ MBConv, e=4, GELU, Layer Norm & 2 & 4 \\
        4 & 96  & 128 & 4 $\times$ MBConv, e=4, GELU, Layer Norm & 2 & 4 \\
        5 & 128 & 256 & 5 $\times$ MBConv, e=4, GELU, Layer Norm & 2 & 4 \\
        \hline
        \multicolumn{6}{|c|}{\textbf{Decoder}} \\
        \hline
        Stage & Chan. In & Chan. Out & Conv. Block & Down-/Up-sampling & Att. Heads \\
        \hline
        1 & 256 & 128 & 4 $\times$ MBConv, e=4, GELU, Layer Norm & 2 & - \\
        2 & 128 & 96  & 3 $\times$ MBConv, e=4, GELU, Layer Norm & 2 & - \\
        3 & 96  & 64  & 2 $\times$ MBConv, e=2, GELU, Layer Norm & 2 & - \\
        4 & 64  & 64  & 2 $\times$ MBConv, e=2, GELU, Layer Norm & 2 & - \\
        \hline
        \multicolumn{6}{|c|}{\textbf{Head}} \\
        \hline
        Stage & Chan. In & Chan. Out & \multicolumn{3}{c|}{Conv. Block} \\
        \hline
        1 & 64  & 32  & \multicolumn{3}{c|}{1 $\times$ Conv2d}  \\
        \hline
    \end{tabular}
\end{table}

\subsection{}     %% Appendix A1, A2, etc.

\noappendix       %% use this to mark the end of the appendix section. Otherwise the figures might be numbered incorrectly (e.g. 10 instead of 1).

%% Regarding figures and tables in appendices, the following two options are possible depending on your general handling of figures and tables in the manuscript environment:

%% Option 1: If you sorted all figures and tables into the sections of the text, please also sort the appendix figures and appendix tables into the respective appendix sections.
%% They will be correctly named automatically.

%% Option 2: If you put all figures after the reference list, please insert appendix tables and figures after the normal tables and figures.
%% To rename them correctly to A1, A2, etc., please add the following commands in front of them:

\appendixfigures  %% needs to be added in front of appendix figures

\appendixtables   %% needs to be added in front of appendix tables

%% Please add \clearpage between each table and/or figure. Further guidelines on figures and tables can be found below.

\authorcontribution{
  SP has implemented the Sim V08 framework and performed the with support from PB. PB and CK have provided feedback throughout the project and contributed to the writing of the manuscript.
} %% this section is mandatory

\competinginterests{No competing interests are present.} %% this section is mandatory even if you declare that no competing interests are present

\disclaimer{TEXT} %% optional section

\begin{acknowledgements}
The work of SP, PB, and CK has been supported by NASA grant
80NSSC22K0604.

GPT-5.6 and GPT-6 were used for copy editing of the manuscript.

\end{acknowledgements}

%% REFERENCES

%% The reference list is compiled as follows:

%\begin{thebibliography}{}
%
%\bibitem[AUTHOR(YEAR)]{LABEL1}
%REFERENCE 1
%
%\bibitem[AUTHOR(YEAR)]{LABEL2}
%REFERENCE 2
%
%\end{thebibliography}

%% Since the Copernicus LaTeX package includes the BibTeX style file copernicus.bst,
%% authors experienced with BibTeX only have to include the following two lines:
%%
\bibliographystyle{copernicus}
\bibliography{references.bib}

\end{document}